\documentclass{book}
\usepackage{amsmath,amsthm}
\usepackage{amstext}
\usepackage{amssymb}
\usepackage{url}

\newtheorem{theorem}{Theorem}[section]

\newtheorem{definition}[theorem]{Definition}
\newtheorem{corollary}[theorem]{Corollary}

\newcommand{\Rt}{\mathbb{R}^3}
\newcommand{\Su}{\mathcal{S}}
\newcommand{\dv}{dv}
\newcommand{\dvf}{dv_0}

\newcommand{\dsf}{ds_0}

\begin{document}

\chapter{Positive energy theorems in General Relativity}

\section{Introduction}
\label{s:intro}

The aim of this chapter is to present an introduction and also
an overview of some of the most relevant results concerning
positivity energy theorems in General Relativity. These
theorems provide the answer to a long standing problem that has
been proved remarkably difficult to solve. They constitute one
of the major results in classical General Relativity and they
uncover a deep self-consistence of the theory.

In this introductory section I would like to present the
theorems in a complete form but with the least possible amount
of technical details, in such a way that the reader can have a
rough idea of the basic ingredients. The examples that
illustrate the hypothesis of the theorems are discussed in the
following sections.

An isolated system is an idealization in physics that assumes
that the sources are confined to a finite region and the fields
are weak far away from the sources. This kind of systems are
expected to have finite total energy. In General Relativity
there are several ways of defining isolated systems. For our
purpose the most appropriate definition is through initial
conditions for Einstein equations. The reasons for that are
twofold. First, the notion of total energy has been discovered
and formulated using a Hamiltonian formulation of the theory
which involves the study of initial conditions. We refer the
reader to the chapter of Domenico Giulini for this topic.
Second, the proofs of the positive mass theorem are mainly
given in terms of initial conditions. For a discussion of the
initial value formulation of Einstein equations we refer to the
chapter of James Isenberg.

Initial conditions for Einstein equations are characterized by
\emph{initial data set} given by $(S, h_{ij}, K_{ij}, \mu,
j^i)$ where $S$ is a connected 3-dimensional manifold, $h_{ij}
$ a (positive definite) Riemannian metric, $K_{ij}$ a symmetric
tensor field, $\mu$ a scalar field and $j^i$ a vector field on
$S$, such that the constraint equations
\begin{align}
 \label{const1}
   D_j   K^{ij} -  D^i   K= -8\pi j^i,\\
 \label{const2}
   R -  K_{ij}   K^{ij}+  K^2=16\pi \mu,
\end{align}
are satisfied on $S$. Here $D$ and $R$ are the Levi-Civita
connection and scalar curvature associated with $ {h}_{ij}$,
and $ K = K_{ij} h^{ij}$. In these equations the indices
$i,k,\ldots$ are 3-dimensional indices, they are raised and
lowered with the metric $ h_{ij}$ and its inverse $ h^{ij}$.
The matter fields are assumed to satisfy the dominant energy
condition
\begin{equation}
  \label{eq:doc}
  \mu\geq \sqrt{j^j j_j}.
\end{equation}

The initial data model an isolated system if the fields are
weak far away from sources. This physical idea is captured in
the following definition of asymptotically flat initial data
set. Let $B_R$ be a ball of finite radius $R$ in $\Rt$. The
exterior region $U=\Rt\setminus B_R$ is called an \emph{end}.
On $U$ we consider Cartesian coordinates $x^i$ with their
associated euclidean radius $r=\left( \sum_{i=1}^3 (x^i)^2
\right)^{1/2}$ and let $\delta_{ij}$ be the euclidean metric
components with respect to $x^i$.  A 3-dimensional manifold $S$
is called \emph{Euclidean at infinity}, if there exists a
compact subset $\mathcal{K}$ of $S$ such that $S\setminus
\mathcal{K}$ is the disjoint union of a finite number of ends
$U_k$.  The initial data set $(S, h_{ij}, K_{ij}, \mu, j^i)$ is
called \emph{asymptotically flat} if $S$ is Euclidean at
infinity and at every end the metric $h_{ij}$ and the tensor
$K_{ij}$ satisfy the following fall off conditions
\begin{equation}
  \label{eq:99}
  h_{ij}=\delta_{ij} +\gamma_{ij}, \quad K_{ij}=O(r^{-2}),
\end{equation}
where $\gamma_{ij}=O(r^{-1})$, $\partial_k \gamma_{ij}=O(r^{-2})$,
$\partial_l\partial_k\gamma_{ij}=O(r^{-3})$ and $\partial_k K_{ij}=O(r^{-3})$.
These conditions are written in terms of Cartesian coordinates $x^i$ attached
at every end $U_k$. Here $\partial_i$ denotes partial  derivatives with respect to these
coordinates.

At first sight it could appear that the notion of asymptotically flat manifold
with ``multiple ends'' $U_k$ is a bit artificial. Certainly, the most important
case is when $S=\Rt$, for which this definition trivializes with
$\mathcal{K}=B_R$ and only one end $U=\Rt\setminus B_R$.  Initial data for
standard configurations of matter like stars or galaxies are modeled with
$S=\Rt$. Also, gravitational collapse can be described with this kind of data.
However, initial conditions with multiple ends and non-trivial interior
$\mathcal{K}$ appear naturally in black hole initial data as we will see. In
particular, the initial data for the Schwarzschild black hole has two
asymptotic ends. On the other hand, this generalization does not imply any
essential difficulty in the proofs of the theorems.

Only conditions on $h_{ij}$ and $K_{ij}$ are imposed in
(\ref{eq:99})  and not on the matter fields $\mu$ and $j^i$,
however since they are coupled by the constraint equations
(\ref{const1})--(\ref{const2}) the fall off conditions
(\ref{eq:99}) impose also fall off conditions on  $(\mu, j^i)$.

The fall off conditions \eqref{eq:99} are far from being the
minimal requirements for the validity of the theorem. This is a
rather delicate issue that have important consequences in the
definition of the energy. We will discuss this point in section
\ref{s:energy}. We have chosen these particular fall off
conditions because they are simple to present and they
encompass a rich family of physical models.

For asymptotically flat initial data the expressions for the total energy and
linear momentum of the spacetime were discovered in \cite{Arnowitt62} and they
are called the ADM energy and linear momentum. They are defined as integrals
over 2-spheres at infinity at every end by the following formulas
\begin{align}
  \label{eq:EP}
  E &=\frac{1}{16\pi}\lim_{r\to \infty} \oint_{\Su_r} \left(\partial_j
    h_{ij}-\partial_i
    h_{jj}\right ) s^i \dsf. \\
  P_i &= \frac{1}{8\pi} \lim_{r\to \infty} \oint_{\Su_r}
  \left(K_{ik}-Kh_{ik} \right ) s^k \dsf, \label{eq:EPP}
\end{align}
where $s^i$ is its exterior unit normal and $\dsf$ is the
surface element of the 2-sphere with respect to the euclidean
metric.   
We emphasize that for every end $U_k$ we have a
corresponding energy and linear momentum $E_{(k)}, P^i_{(k)}$,
which can have different values. We will discuss examples of
that in section \ref{s:energy}.

The quantities $E$ and $P_i$ are defined on the asymptotic ends
and they depend only on the asymptotic behaviour of the fields
$h_{ij}$ and $K_{ij}$. However, since $h_{ij}$ and $K_{ij}$
satisfy the constraint equations \eqref{const1}--\eqref{const2}
and the dominant energy condition \eqref{eq:doc} holds these
quantities carry in fact information of the whole initial
conditions.

The energy $E$ and the linear momentum $P_i$ are components of
a 4-vector $P_a=(E,P_i)$ (indices $a,b,c,\ldots$ are
4-dimensional). We will discuss this in section
\ref{s:momentum}.  The total mass of the spacetime is defined
by
\begin{equation}
  \label{eq:86}
  M=\sqrt{E^2-P_i P_j\delta^{ij}}.
\end{equation}

We have all the ingredients to present the positive energy theorem.
\begin{theorem}[Positive energy theorem]
\label{t:pmt}
  Let $(S,h_{ij},K_{ij},\mu, j^i)$ be an asymptotically flat (with possible many
  asymptotic ends), complete,  initial data set, such that
  the dominant energy condition (\ref{eq:doc}) holds. Then  the energy and
  linear momentum
  $(E,P_i)$ defined by (\ref{eq:EP})--\eqref{eq:EPP} satisfies
  \begin{equation}
    \label{eq:100}
    E\geq \sqrt{ P_i P_j\delta^{ij} }\geq 0.
  \end{equation}
  at every end. Moreover, $E=0$ at any end if and only if the initial data
  correspond to the Minkowski
  space-time.
\end{theorem}
The word ``complete'' means that $(S,h_{ij})$ as Riemannian manifold is
complete. That is, no singularities are present on the initial conditions.  But
the space-time can be singular since singularities can developed from regular
initial conditions, for example in the gravitational collapse. We will discuss
that in more detail in section \ref{s:energy}.

Note that Theorem \ref{t:pmt} allows the vector $P^a$ to be null and non
trivial. However, it has been shown in \cite{Ashtekar-horowitz82} that if the
energy momentum vector $P^a$ is null then it vanishes identically. In
\cite{Beig96d} this result has been proved without imposing conditions on
fields other than $h_{ij}$ and $K_{ij}$.

One remarkable aspect of this theorem is that it is non-trivial
even in the case where $S=\Rt$ and no matter fields $\mu=j^i=0$
are present. This correspond to the positivity of the energy of
the pure vacuum gravitational waves. We present explicit
examples of this in section \ref{s:energy}.

For spacetimes with black holes there are spacelike surfaces that touch the
singularity. For that kind of initial conditions theorem \ref{t:pmt} does not
apply. Physically it is expected that it should be possible to prove a
positivity energy theorem for black holes without assuming anything about what
happens inside the black hole.  That is, it should be possible to prove an
extension of the positive energy theorem for initial conditions with inner
boundaries if the boundary represents a black hole horizon. The following
theorem deals precisely with that problem.

\begin{theorem}[Positive energy theorem with black hole inner boundaries]
\label{t:pmtbh}
  Let $(S,h_{ij},K_{ij})$ be an asymptotically flat, complete,  initial data set, with
  $S=\Rt\setminus B$, where $B$ is a ball. Assume  that
  the dominant energy condition \eqref{eq:doc} holds and and  that $\partial B$ is a black
  hole boundary.  Then  the energy momentum
  $E,P^i$ defined by (\ref{eq:EP})--~\eqref{eq:EPP}  satisfies
  \begin{equation}
    \label{eq:100b}
    E\geq \sqrt{P^iP_i}\geq 0.
  \end{equation}
Moreover, $E=0$  if and only if the initial data correspond to the Minkowski space-time.
\end{theorem}
We will explain what are black hole inner boundary conditions in section
\ref{s:energy}.

The plan of the chapter is the following. In section \ref{s:energy} we discuss
the concept of the energy $E$ and we  present examples that illustrate the
hypothesis of the positive energy theorem. In section \ref{s:momentum} we
analyze the linear momentum $P_i$ and describe its transformation properties.
In section \ref{s:proof} we review the main steps of the proof of theorems
\ref{t:pmt} and \ref{t:pmtbh}. Finally in section \ref{s:gi} other recent
related results are discussed and the relevant current open problems are presented.

\section{Energy}
\label{s:energy} A remarkable feature of the asymptotic
conditions \eqref{eq:EP} is that they imply that the total
energy can be expressed exclusively in terms of the Riemannian
metric $h_{ij}$ of the initial data (and the linear momentum in
terms of $h_{ij}$ and the second fundamental form $K_{ij}$).
Hence the notion of energy can be discussed in a pure
Riemannian setting, without mention the second fundamental
form. Moreover, as we will see, there is a natural corollary of
the positive energy theorem for Riemannian manifolds.  This
corollary is relevant for several reasons. First, it provides a
simpler and relevant setting to prove the positive energy
theorem. Second, and more important, it has surprising
applications in other areas of mathematics.  Finally, to deal
first with the Riemannian metric and then, in the next section,
with the second fundamental form to incorporate the linear
momentum, reveal the different mathematical structures behind
the energy concept.

In the previous section we have introduced the notion of an end $U$, the energy
is defined in terms of Riemannian metrics on $U$.  To emphasize this important
point we isolate the notion of energy defined in the introduction in the
following definition.
\begin{definition}[Energy]
\label{d:mass}
Let  $h_{ij}$ be a Riemannian metric on an end $U$ given in
the coordinate system $x^i$ associated with $U$. The energy is defined by
\begin{equation}
  \label{eq:30adm}
  E=\frac{1}{16\pi}\lim_{r\to \infty} \oint_{\Su_r} \left(\partial_j
    h_{ij}-\partial_i
    h_{jj}\right ) s^i \dsf.
\end{equation}
\end{definition}
Note that in this definition there is no mention to the constraint equations
\eqref{const1}--\eqref{const2}. Also, the definition only involve an end $U$,
there is no assumptions on the interior of the manifold.

In the literature it is custom to call $E$ the total mass and denote it by $m$
or $M$. In this article, in order to emphasize that $E$ is in fact the zero
component of a four vector we prefer to call it energy and reserve the name
mass to the quantity $M$ defined by (\ref{eq:86}).  When the linear momentum is
zero, both quantities coincides.

The definition of the total energy has three main ingredients: the end $U$, the
coordinate system $x^i$ and the Riemannian metric $h_{ij}$. The metric is
always assumed to be smooth on $U$, we will deal with singular metrics but
these singularities will be in the interior region of the manifold and not
on $U$.

There exists two potentials problems with the definition \ref{d:mass}. The
first one is that the integral (\ref{eq:30adm}) could be infinite. The second,
and more subtle, problem is that the mass seems to depend on the particular
coordinate system $x^i$. Both problems are related with fall off conditions for
the metric.  In the previous section we have introduced in equation
\eqref{eq:99} an example of this kind of conditions. These conditions are
probably sufficient to model most physically relevant initial
data. However, it is interesting to study the optimal fall off conditions
that are necessary to have a well defined notion of energy  and such that the
energy is independent of the coordinate system.

To study this problem, we introduce first a general class of fall off conditions
as follows.  Given an end $U$ with coordinates $x^i$, and an arbitrary real
number $\alpha$, we say that the metric $h_{ij}$ on $U$ is \emph{asymptotically
  flat of degree} $\alpha$ if the components of the metric with respect to
these coordinates have the following fall off in $U$ as $r\to \infty$
\begin{equation}
  \label{eq:54}
  h_{ij}=\delta_{ij}+\gamma_{ij},
\end{equation}
with $\gamma_{ij}=O(r^{-\alpha})$, $\partial_k
\gamma_{ij}=O(r^{-\alpha-1})$. The subtle point is to determine
the appropriate $\alpha$ decay. To understand the meaning of
this coefficient let us discuss the following relevant example
given in \cite{denisov83} (see also \cite{Bray04}). Take the
euclidean metric $\delta_{ij}$ in Cartesian coordinates $x^i$
and consider coordinates $y^i$ defined by
\begin{equation}
  \label{eq:4}
  y^i=\frac{\rho}{r}x^i, \quad
\end{equation}
where $\rho$ is defined by
\begin{equation}
  \label{eq:55}
  r=\rho+c\rho^{1-\alpha},
\end{equation}
for some constants $c$ and $\alpha$. Note that
$\rho=\left( \sum_{i=1}^3 (y^i)^2 \right)^{1/2}$. The components $g'_{ij}$ of
the euclidean metric in coordinates $y^i$ have the following form
\begin{equation}
  \label{eq:19}
  g'_{ij}=\delta_{ij} + \gamma_{ij},
\end{equation}
where $\gamma_{ij}$ satisfies the decay conditions \eqref{eq:54} with the
arbitrary $\alpha$ prescribed in the coordinate definition \eqref{eq:55}. That
is, the metric in the new coordinate system $y^i$ is asymptotically flat of
degree $\alpha$.

We calculate the energy in the coordinates $y^i$ using the definition
(\ref{eq:30adm}). We obtain
\begin{equation}
  \label{eq:20}
  E=\begin{cases} \infty, \quad \alpha < 1/2,\\
c^2/8, \quad \alpha = 1/2,  \\
0,  \quad \alpha >1/2.
 \end{cases}
\end{equation}
Of course, we expect that the energy of the euclidean metric should be zero in
any coordinate system.  The interesting point of this example is the limit case
$\alpha=1/2$, the example shows that if the energy has any chance to be
coordinate independent, then we should impose $\alpha >1/2$.  The following
theorem, proved in \cite{Bartnik86} and \cite{chrusciel86}, says that this
condition is also sufficient (see also \cite{murchadha:2111} where optimal fall
off conditions are analyzed also for the linear momentum).

\begin{theorem}
\label{t:um}
Let $U$ be an end with a Riemannian metric $h_{ij}$ such that is satisfies the
fall off conditions \eqref{eq:54} with $\alpha >1/2$.
Assume also that the scalar curvature $R$ is integrable in $U$, that is
\begin{equation}
  \label{eq:12}
\int_U  |R|\, \dv <\infty.
\end{equation}
Then the energy defined by \eqref{eq:30adm} is unique and it is finite.
\end{theorem}
In this theorem unique means if we calculate the energy in any
coordinate system for which the metric satisfies the decay
conditions \eqref{eq:54} with $\alpha >1/2$ we obtain the same
result.  This theorem ensure that the energy is a geometrical
invariant of the Riemannian metric in the end $U$.
Historically, this theorem was proved after the positive energy
theorems. In the original proofs of the positive energy
theorems different decay conditions for the metric have been
used.  The decay conditions are usually formulated in terms of
integrals of derivatives (i.e. Sobolev spaces) (see
\cite{Bartnik86}) which are more flexible for many
applications. This particular formulation (which is simpler to
present) of theorem \ref{t:um} was taken from
\cite{chrusciel12}.  The decay conditions with $\alpha >1/2$
together with the condition \eqref{eq:12} on the scalar
curvature are called \emph{mass decay
  conditions}.  The freedom in the coordinates $x^i$ is only a rigid motion at
infinity (see \cite{Bartnik86}).

Theorem \ref{t:um} completes the geometric characterization of the energy at the end
$U$. We turn now to positivity. It is clear that the energy can have any sign
on $U$.  The model example is given by the initial data for the Schwarzschild
black hole, with metric on $U$ given by
\begin{equation}
  \label{eq:3}
  h_{ij}=\psi^4 \delta_{ij},
\end{equation}
where $\psi$ is the following function
\begin{equation}
  \label{eq:1}
  \psi=1+\frac{C}{2r},
\end{equation}
with $C$  an arbitrary constant. Computing the energy for this metric we
obtain $E=C$. The constant $C$ can of course have any sign. It is however
important to emphasize that theorem \ref{t:um} asserts that the energy is well
defined and it is an invariant of the geometry of the end even when it is
negative.

To ensure the positivity of the energy we need to impose two
important conditions. One is a local condition: the positivity
of the local energy given by the dominant energy condition
(\ref{eq:doc}). The other is a global condition on the
manifold: the manifold should be complete or should have black
hole boundaries.

Initial conditions with
\begin{equation}
  \label{eq:89}
  K_{ij}=0,
\end{equation}
are called time symmetric initial data. That is, time symmetric
initial data are characterized only by a Riemannian metric
$h_{ij}$. Conversely, any Riemannian metric can be interpreted
as a time symmetric initial data. However, an arbitrary metric
will not satisfy the dominant energy condition (\ref{eq:doc}).
In effect, inserting condition \eqref{eq:89} in the constraint
equation (\ref{const2}) and using  the dominant energy
condition (\ref{eq:1}) we obtain
\begin{equation}
  \label{eq:80}
  R \geq 0.
\end{equation}
Only metrics that satisfy \eqref{eq:80} can be interpreted as time
symmetric initial data for which the dominant energy condition holds. But then,
any metric such that \eqref{eq:80} holds satisfies the dominant energy
condition and it is a good candidate for the positive energy theorem. And hence
we obtain the following corollary of theorem \ref{t:pmt}.

\begin{corollary}[Riemannian positive mass theorem]
\label{t:rpmt}
Let $(S,h_{ij})$ be a complete, asymptotically flat, Riemannian manifold.
Assume that the scalar curvature is non-negative (i.e. condition
\eqref{eq:80}). Then the energy is non-negative at every end and it is zero at
one end if an only if the metric is flat.
\end{corollary}
This corollary was proved with the optimal decay conditions for the metric in
\cite{Bartnik86} and \cite{Lee87}.

The interesting mathematical aspect of this corollary is that there is no
mention to the constraint equations, the second fundamental form or the matter
fields. This theorem is a result in pure Riemannian geometry. It has surprising
applications in the solution of the Yamabe problem (see the review article
\cite{Lee87} and reference therein).

Note that it is not necessary to impose that the whole second fundamental form
is zero to have (\ref{eq:80}), from equation \eqref{const2} it is clear that is
enough to have $K=0$. This class of initial data are called maximal and they
have important properties (see the chapter by J. Isenberg). In particular,
positive energy theorems for this kind of data are easier to prove (mainly
because condition \eqref{eq:80} holds) than for general initial data.

Let us discuss some examples of corollary \ref{t:rpmt}.  We begin with the case
with one asymptotic end and trivial topology, namely $S=\Rt$. For arbitrary
functions $\psi$, metrics of the form \eqref{eq:3} are called conformally flat,
they provide a very rich family of initial conditions which have many
interesting applications (for example, initial data for black hole collisions,
see the review article \cite{Cook00}). The scalar curvature for this class of
metrics is given by
\begin{equation}
  \label{eq:15}
  R=-8\psi^{-5} \Delta \psi,
\end{equation}
where $\Delta$ is the euclidean Laplacian. If $\psi$ satisfies the fall off conditions
\begin{equation}
  \label{eq:13b}
  \psi=1+u,\quad   u=O(r^{-1}), \quad  \partial_k u =O(r^{-2}),
\end{equation}
then the energy for this class of metric is given by
\begin{equation}
  \label{eq:14}
  E=-\frac{1}{2\pi} \lim_{r\to \infty} \oint_{\Su_r} \partial_r \psi \, \dsf.
\end{equation}

For $\psi$  given by \eqref{eq:1} we obtain $R=0$, and then the metric satisfies the
local condition \eqref{eq:80} for any choice of the constant $C$. However,
this metric can not be extended to $\Rt$ since the function $\psi$ is singular
at $r=0$ and hence, as expected, corollary \ref{t:rpmt} does not apply to this
case. Let us try to prescribe a function with the same decay (and hence
identical energy) but such that it is regular at $r=0$.
For example
\begin{equation}
  \label{eq:5}
   \psi=1+\frac{C}{2\sqrt{r^2 +C^2}}.
\end{equation}
Using  (\ref{eq:14}) we obtain again that $E=C$. For any value of $C$ the
function $\psi$ is strictly positive and bounded on $\Rt$ and hence the metric
is smooth on $\Rt$. That is, it satisfies the completeness assumption in
corollary \ref{t:rpmt}. Using \eqref{eq:15} we compute the scalar curvature
\begin{equation}
  \label{eq:16}
  R = 12 \psi^{-5} \frac{C^3}{(r^2+C^2)^{5/2}}.
\end{equation}
We have $R\geq 0$ if and only if $C\geq 0$. Also, in this example the
mass is zero if and only if the metric is flat.

Other  interesting examples can be constructed with conformally flat
metrics as follows.  Let $\psi$ be a solution of the Poisson equation
\begin{equation}
  \label{eq:87}
  \Delta \psi=-2\pi \tilde \mu,
\end{equation}
that satisfies the decay conditions \eqref{eq:13b}, where $\tilde \mu$ is
a non-negative function of compact support in $\Rt$. Solution of (\ref{eq:87})
can be easily constructed using the Green function of the Laplacian. By
equation \eqref{eq:15}, the scalar curvature of the associated conformal metric
\eqref{eq:3} will be non-negative and the function $\tilde \mu$ is related to
the matter density $\mu$ by
\begin{equation}
  \label{eq:88}
  \mu=\frac{R}{16\pi}= \tilde \mu \psi^{-5}.
\end{equation}
Note that we can not prescribe, in this example, exactly the
matter density $\mu$, we prescribe a conformal rescaling of
$\mu$. However, it is enough to control de support of $\mu$.
The support of $\mu$ represents the localization of the matter
sources. Outside the matter sources the scalar curvature (for
time symmetric data) is zero.

For conformally flat metrics in $\Rt$ there is a very simple proof of
corollary \ref{t:rpmt}. We write equation \eqref{eq:15} as
\begin{equation}
  \label{eq:13}
  \frac{R}{8}=-\partial^i\left( \frac{\partial_i \psi}{\psi^5} \right) -5
  \frac{|\partial \psi|^2}{\psi^6}.
\end{equation}
Integrating this equation in $\Rt$, using for the first term in the
right-hand side the Gauss theorem, the condition $\psi \to 1$ as $r \to \infty$
and the expression \eqref{eq:14} for the energy we finally obtain
\begin{equation}
  \label{eq:21}
  E=\frac{1}{2\pi} \int_{\Rt} \left( \frac{R}{8} + 5
  \frac{|\partial \psi|^2}{\psi^6} \right) \, \dvf,
\end{equation}
where $\dvf$ is the flat volume element.
This formula proves that for metric of the form \eqref{eq:3} we have $E\geq 0$
if $R\geq 0$ and $E=0$ if and only if $h_{ij}=\delta_{ij}$. This proof easily
generalize for conformally flat maximal initial data.

Asymptotically flat initial conditions in $\Rt$ with no matter sources
(i.e. $\mu=j^i=0$) represent pure gravitational waves. They are conceptually
important because they describe the dynamic of pure vacuum, independent of any
matter model.  Note that in that case the energy condition \eqref{eq:doc} is
trivially satisfied.

In the previous examples the only solution with pure vacuum $R=0$ in $\Rt$ is
the flat metric, because by equation \eqref{eq:15} we obtain $\Delta\psi=0$ and
the decay condition \eqref{eq:13b} implies $\psi=1$.  In order to construct
pure waves initial data  we allow for more general kind of conformal metrics,
let $h_{ij}$ be given by
\begin{equation}
  \label{eq:90}
   h=  e^\sigma\left[e^{-2q}(d\rho^2+dz^2)+ \rho^2 d\varphi^2\right],
\end{equation}
where $(\rho, z, \varphi)$ are cylindrical coordinates in $\Rt$ and the
functions $q$ and $\sigma$ depend only on $(\rho,z)$. That is, the metric
$h_{ij}$ given by \eqref{eq:90} is axially symmetric.

The scalar curvature of the metric \eqref{eq:90} is given by
\begin{equation}
  \label{eq:92}
  -\frac{1}{8}  R e^{(\sigma-2q)} = \frac{1}{4}\Delta \sigma +\frac{1}{16}
|\partial \sigma|^2
- \frac{1}{4}\Delta_2 q ,
\end{equation}
where $\Delta$, as before, is the 3-dimensional flat Laplacian and $\Delta_2$
is the 2-dimensional Laplacian in cylindrical coordinates given by
\begin{equation}
  \label{eq:22}
  \Delta_2 q=\partial^2_\rho q+\partial^2_z q.
\end{equation}
If we impose $R=0$, equation \eqref{eq:92} reduce to
\begin{equation}
  \label{eq:102}
  \Delta \psi -\frac{1}{4} \Delta_2 q =0,
\end{equation}
where $\psi^4=e^\sigma$. To construct metrics  of the form \eqref{eq:90} that satisfies
$R=0$ a function $q$ is prescribed and then the linear equation \eqref{eq:102}
is solved for $\psi$. The function $q$ can not be arbitrary, it should satisfy
a global condition (which is related with the Yamabe problem mentioned above),
see \cite{Cantor81b} for details.
This kind of metric are called Brill waves, they have been used by
D. Brill in one of the first proofs of the positive energy theorem
\cite{Brill59}. Let us discuss this proof.

In order to be smooth at the axis the metric \eqref{eq:90}  should satisfies
$q=0$ at $\rho=0$. For simplicity we also impose a strong fall off condition on
$q$ at infinity, namely
$q=O(r^{-2})$, $\partial_i q=O(r^{-2})$. For $\sigma$ we impose
$\sigma=O(r^{-1})$ and $\partial_i \sigma =O(r^{-2})$. Using these decay
assumptions is straightforward to check that the energy of the metric
\eqref{eq:90} is given by
\begin{equation}
  \label{eq:101}
  E=-\frac{1}{8\pi} \lim_{r\to \infty} \oint_{\Su_r} \partial_r \sigma \, \dsf.
\end{equation}

By Gauss theorem, using that $q=0$ at the axis and the fall off condition of
$q$ at infinity we obtain that
\begin{equation}
  \label{eq:93}
  \int_{\Rt} \Delta_2 q \, \dvf=0.
\end{equation}

Integrating equation \eqref{eq:92} in $\Rt$, using \eqref{eq:93} and
using the expression \eqref{eq:101} for the energy we obtain
\begin{equation}
  \label{eq:94}
  E= \frac{1}{8\pi} \int_{\Rt} \left(\frac{1}{2}
|\partial \sigma|^2 + Re^{\sigma-2q}\right) \,\dvf.
\end{equation}
That is, $R\geq 0$ implies $E\geq 0$.
In particular for  vacuum $R=0$, we have
\begin{equation}
  \label{eq:95}
  E= \frac{1}{16\pi}\int_{\Rt} | \partial \sigma|^2  \,\dvf.
\end{equation}
This positivity proof can  be extended in many ways, for example to maximal
initial data \cite{Murchadha74}.  In particular it has
applications for the inequality between energy and angular momentum discussed
in section \ref{s:gi} (see the review article \cite{dain12} and the lectures notes
\cite{chrusciel08} \cite{chrusciel12}, and reference therein).

We turn now to manifolds with many asymptotic flat ends and  interior
$\mathcal{K}$ with non-trivial topology defined in section \ref{s:intro}. Let
us first present some basic example of the definition of asymptotic euclidean
manifold, without mention the metric.

Take out a point in $\Rt$, the manifold $S=\Rt\setminus \{0\}$ is asymptotic
Euclidean with two ends, which we denote by $U_0$ and $U_1$. In effect, let
$B_2$ and $B_1$ be two balls centered at the origin with radius $2$ and $1$
respectively. Define $\mathcal{K}$ be the annulus centered
at the origin $B_2\setminus B_1$.  Then $S\setminus \mathcal{K} $ has two
components $U_0$ and $U_1$, where $U_0=\Rt\setminus B_2$ and $U_1=B_1 \setminus
\{0\} $. The set $U_0$ is clearly an end. The set $U_1$ is also an end since
the a ball minus a point is diffeomorphic to $\Rt$ minus a ball. This can be
explicitly seen using Cartesian coordinates centered at the origin $x^i$, then
the map given by the inversion
\begin{equation}
  \label{eq:2}
  y^i =r^{-2} x^i,
\end{equation}
provide the diffeomorphism between $\Rt\setminus B_1$ and $B_1\setminus
\{0\}$.

In the same way $\Rt$ minus a finite number $N$ of points $i_k$ is an Euclidean
manifold with $N+1$ ends.  For each $i_k$ take a small ball $B_k$ of
radius $r_{(k)}$, centered at $i_k$, where $r_{(k)}$ is small enough such that
$B_k$ does not contain any other $i_{k'}$ with $k'\neq k$. Take $B_R$, with
large $R$, such that $B_R$ contains all points $i_k$. The compact set
$\mathcal{K}$ is given by $\mathcal{K}= B_R \setminus \sum_{k=1}^N B_k$ and the
open sets $U_k$ are given by $B_k\setminus i_k$, for $1 \leq k \leq N$, and
$U_0$ is given by $\Rt \setminus B_R$.

Another example is a torus $\mathbb{T}^3$ minus a point $i_0$. Take a small
ball $B$ centered at $i_0$. Then the manifold is asymptotic euclidean with
$\mathcal{K}= \mathbb{T}^3\setminus B$ and only one end $U=B\setminus
i_0$. This is an example of an Euclidean manifold with one asymptotic end but
non-trivial $\mathcal{K}$.
More generally, given any compact manifold, if we subtract a finite
number of points we get an asymptotically Euclidean manifold with multiple
ends. Note that the topology of the compact core $\mathcal{K} $ can be very
complicated.

Let us consider now Riemannian metrics on these asymptotic euclidean manifolds.
Consider the manifold $S=\Rt \setminus \{0\}$ and the metric given by
\eqref{eq:3} and  \eqref{eq:1}.
The function $\psi$ is smooth on $S$ for any value of the constant $C$, however
if $C<0$ then $\psi$ vanished at $r=-2/C$ and hence the metric is not defined
at those points. That is, the metric $h_{ij}$ is smooth on $S$ only when $C\geq
0$. We have seen that $S$ has two asymptotic ends, let us check that the metric
$h_{ij}$ is asymptotically flat (i.e. it satisfies the decay conditions
\eqref{eq:99}) at both ends $U_0$ and $U_1$. On $U_0$, the metric in the
coordinates $x^i$ is clearly asymptotically flat. But note that in this
coordinates the metric is not asymptotically flat at the end $U_1$ (which, in
these coordinates is represented by a neighborhood of $r=0$), in fact the
components of the metric are singular at $r=0$. However, using a coordinate
an inversion of coordinates like \eqref{eq:2} is straightforward  to prove
that the metric is  asymptotically flat also at $r=0$. More precisely, consider
the coordinate transformation
\begin{equation}
  \label{eq:31}
  y^i=\left(\frac{C}{2} \right)^2 \frac{1}{r^2} x^i, \quad
\rho=\left(\frac{C}{2}
  \right)^2 \frac{1}{r}.
\end{equation}
In terms of this coordinates the metric has the form
\begin{equation}
  \label{eq:32}
  h'_{ij}=\left( 1+\frac{C}{2\rho} \right)^4 \delta_{ij}.
\end{equation}
We have chosen the constant factor in the coordinate transformation
\eqref{eq:32} in such a way that the transformation it is in fact the well
known isometry of this metric, this choice is however not essential.  The
metric \eqref{eq:32} is clearly asymptotically flat at the $U_1$.  Note that we
have two energies, one for each end, the two are equal and given by the
constant $C$.  In this example the positivity of the mass is enforced purely by
the global requirement of completeness of the metric (the energy condition is
satisfied for arbitrary $C$). It is this condition that fails when $C<0$.  In that
case the metric is defined on a manifold with boundary $S=\Rt\setminus
B_{-2/C}$, and the metric vanished at the boundary $\partial B_{-2/C}$. In
particular, the 2-surface $\partial B_{-2/C}$ has zero area. This motivated the
concept of ``zero area singularities'' introduced in \cite{Bray:2009cw}, where
interesting results are presented concerning negative energy defined on this
class of singular metrics.

In the previous example the energies at the different ends are equal.
It is straightforward to construct an example for which the two energies are
different. Consider the following function
\begin{equation}
  \label{eq:33}
  \psi= 1+ \frac{C}{2r}+g,
\end{equation}
where $g$ is a smooth function on $\Rt$ such that $g=O(r^{-2})$ as $r\to
\infty$ and $g(0)=a$. Making the same calculation we get that the energy at one
end is $E_0=C$ (here we use the decay conditions on $g$, otherwise the function
$g$ will contribute to the energy at that end). But at the other end the components of the
metric in the coordinates $y^i$ are given by
\begin{equation}
  \label{eq:34}
   h'_{ij}=\left( 1+\frac{C(1+g)}{2\rho} \right)^4 \delta_{ij},
\end{equation}
and hence we have that
\begin{equation}
  \label{eq:35}
  E_1=C(1+a).
\end{equation}
Note that in order to satisfy the energy condition \eqref{eq:80}  $g$  (and
hence $a$) can not be arbitrary, we must
impose the following condition on $g$
\begin{equation}
  \label{eq:56}
  \Delta g \leq 0.
\end{equation}
Using \eqref{eq:56}, the decay assumption on $g$ and the maximum principle for
the Laplacian (see, for example, the version of the maximum principle in the
appendix of \cite{Dain:2008yu}) it is easy to prove that $g\geq 0$ and then $a\geq 0$.

Consider the manifold $S=\Rt\setminus \{i_1\}, \{i_2\}$ with three asymptotic
ends. And consider the function given by (this nice example was constructed in
\cite{Brill63})
\begin{equation}
  \label{eq:7}
  \psi= 1+ \frac{C_1}{2r_1}+\frac{C_2}{2r_2}.
\end{equation}
where $r_1$ and $r_2$ are the euclidean radius centered at the points $i_1$ and
$i_2$ respectively, and  $C_1$ and $C_2$ are constant. Note that $\Delta
\psi=0$ and hence the metric defined by \eqref{eq:3} has $R=0$. As before, only
when $C_1,C_2\geq 0$ the metric is smooth on $S$. Also,
using a similar calculation as in the case of two ends it is not difficult to
check that the metric is asymptotically flat on the three ends. Moreover, the
energies of the different ends are given by
\begin{equation}
  \label{eq:37}
  E_0=C_1+C_2, \quad E_1=C_1+\frac{C_1C_2}{L}, \quad  E_2=C_2+\frac{C_1C_2}{L},
\end{equation}
where $L$ be the euclidean distance between $i_1$ and $i_2$. We see that they
are all positive and, in general, different.
These initial conditions model a head on collision of two black holes and they
have been extensively used in numerical simulations of black hole collisions
(see for example \cite{alcubierre-it3nrsomop2008} and reference therein).

We analyze the case of the  wormhole (see \cite{Misner60}), which is an example
of a $\mathcal{K}$ with more complicated topology.
Consider the  metric on the compact manifold $S=\mathbb{S}^1\times\mathbb{S}^2$
given by
\begin{equation}
  \label{eq:8}
  \gamma=d\mu^2 +(d\theta^2 +\sin^2 \theta d\varphi^2),
\end{equation}
where the coordinates ranges are $-\pi <\mu\leq \pi$, and the sections
$\mu=const$ are 2-spheres.
Let  $h_{ij}$ be given by
\begin{equation}
  \label{eq:38}
  h_{ij}=\psi^4 \gamma_{ij}
\end{equation}
where the function $\psi$ is
\begin{equation}
  \label{eq:39}
  \psi=\sum_{n=-\infty}^{n=\infty} \left[ cosh(\mu+2n\pi \right)]^{-1/2}.
\end{equation}
This function blows up at $\mu=0$. Hence, the metric $h_{ij}$ is defined on
$S$ minus the point $\mu=0$. We have seen that this is an asymptotic euclidean
manifold with one asymptotic end. It can be proved that the metric is
asymptotically flat at that end (see \cite{Misner60} for details). Also, the
function $\psi$ is chosen in such a way that the scalar of $h_{ij}$ curvature
vanished.  Moreover, the energy is given by
\begin{equation}
  \label{eq:9}
  E= 4 \sum_{n=1}^\infty (\sinh(n\pi))^{-1},
\end{equation}
which is positive.

Finally, consider the initial data for the  Reissner-N\"ordstrom black hole
given by a metric of the form \eqref{eq:3} with $\psi$ given by
\begin{equation}
  \label{eq:10}
  \psi=\frac{1}{2r} \sqrt{(q+2r+C)(-q+2r+C)},
\end{equation}
where $C$ and $q$ are constant. The scalar curvature of this metric is given by
\begin{equation}
  \label{eq:104}
  R=\frac{2q^2}{\psi^8r^4}.
\end{equation}
Which is non-negative for any value of the constants.
When $C>|q|$, then the metric is asymptotically flat with two ends $U_0$ and
$U_1$ as in the example (\ref{eq:1}) . The energy
on both ends is given by $E=C$. The positive energy theorem applies to this
case. If    $C<|q|$ then the metric is singular, there is only one end $U_0$
and the energy on that end is given by $C$. Note that in this case it is still
possible to have positive energy $0<C<|q|$, but the positive energy theorem does
not apply because is a singular metric. The borderline case $C=|q|$ represent
the extreme black hole. The manifold is $\Rt$ minus a point and the metric is
smooth on that manifold. However the metric is asymptotically flat only at the
end $U_0$, on the other end is asymptotically cylindrical. And hence this
version of the positive energy theorem does not apply for these data. The
asymptotically cylindrical end is a feature of all extreme black holes. For
discussions on this kind of geometry see \cite{dain12} and reference therein.

So far, we have discussed complete manifolds without boundaries or manifolds
with boundaries in which the metric is singular at the boundaries. We analyze
now the important case of black hole boundaries.

Black hole boundaries are defined in terms of marginally trapped surfaces. A
marginally trapped surface is a closed 2-surface such that the outgoing null
expansion $\Theta_+$ vanishes (more details on this important  concept can be
seen in \cite{Wald84}).  If such surface is embedded on a space-like
3-dimensional surface, then the expansion $\Theta_+$ can be written in terms of
the initial conditions as follows
\begin{equation}
  \label{eq:96}
  \Theta_+=H -K_{ij} s^is^j+ K,
\end{equation}
where
\begin{equation}
  \label{eq:97}
  H=D_is^i,
\end{equation}
is the mean curvature of the surface. Here $s^i$ is the unit normal vector to
the surface.
For time symmetric initial data, condition $\Theta_+=0$ reduces to
\begin{equation}
  \label{eq:98}
  H=0.
\end{equation}
Surfaces that satisfies condition (\ref{eq:98}) are called minimal surfaces,
because (\ref{eq:98}) is satisfied if and only if the first variation of the
area of the surface vanishes. These kind of surfaces have been extensively
studied in Riemannian geometry (see the  book \cite{osserman86}  for an
introduction to the subject). We have seen that a marginally trapped surface on
a time symmetric initial data is a minimal surface. That is, black hole
boundaries translate, for these kind of data, into a pure Riemannian boundary
condition. Then, we have the following corollary of theorem \ref{t:pmtbh}.

\begin{corollary}[Black holes in Riemannian geometry]
\label{t:bhrpm}
  Let $(S,h_{ij})$ be a complete, asymptotically flat, Riemannian manifold with
  compact boundary.  Assume that the scalar curvature is non-negative
  (i.e. condition \eqref{eq:80}) and that the boundary is a minimal surface
  (i.e. it satisfies (\ref{eq:98}). Then the energy is non-negative and it is
  zero at one end if and only if the metric is flat.
\end{corollary}
Let as give a very simple example that illustrate this theorem.
Consider the function $\psi$ given by (\ref{eq:1}). It is well known that the
surface $r=C/2$ is a minimal surface (it represents the intersection of the
Schwarzschild black hole event horizon with the spacelike surface $t=constant$
in Schwarzschild coordinates). To verify, that we compute  $H$ for the 2-surfaces
$r=constant$ for the metric (\ref{eq:3}). The unit normal vector is given  by
\begin{equation}
  \label{eq:106}
  s^i= \psi^{-2} \left(\frac{\partial}{\partial r}\right)^i.
\end{equation}
Then we have
\begin{equation}
  \label{eq:105}
  H=D_i s^i= \frac{4}{\psi^{3}}\left( \partial_r\psi+ \frac{\psi}{2r}\right).
\end{equation}
Then. condition (\ref{eq:98}) is equivalent to
\begin{equation}
  \label{eq:107}
 0= \partial_r\psi+ \frac{\psi}{2r}=  \frac{1}{2r}-\frac{C}{4r^2},
\end{equation}
and hence for $r=C/2$ we have a minimal surface. Note that $C$ must be positive
in order to have a minimal surface. Previously we have discussed this example
in the complete manifold, without boundaries, $\Rt\setminus \{0\}$. In that case
corollary \ref{t:rpmt} applies. We can also consider the same metric but in the
manifold with boundary $Rt\setminus B_{C/2}$. Since we have seen that $\partial
B_{C/2}$ is a minimal surface, then corollary \ref{t:bhrpm} applies to that
case. To emphasize the scope of this corollary, we slightly extend this example
in the following form. Consider $\psi$ given by
\begin{equation}
  \label{eq:25}
  \psi=(1+\frac{C}{2r})\chi(r),
\end{equation}
where $\chi(r)$ is a  function such that is $\chi=1$ for $r>C/2$
and arbitrary for $r<C/2$. Corollary \ref{t:bhrpm} applies to this case since again the
boundary is a minimal surface. Note that inside the minimal surface the
function $\chi$ is arbitrary, in particular it can blows up and it does not
need to satisfies the energy condition.  The corollary \ref{t:rpmt} certainly
does not apply to this case.

\section{Linear momentum}
\label{s:momentum}
The total mass $M$ defined by (\ref{eq:86}) in terms of the energy and linear
momentum (\ref{eq:EP})--\eqref{eq:EPP} represents the total amount of energy of
the space-time. The first basic question we need to address is in what sense $M$
is independent of the choice of  initial conditions that describe the same
space-time. That is, given a fixed space-time we can take different space-like
surfaces on it, on each surface we can calculate the initial data set and hence
we have a corresponding $M$, do we get the same result? We will see
that the answer of that question strongly depend on the fall off conditions
(\ref{eq:99}).

To illustrate that, let us consider the Schwarzschild space-time. We recall that
in the following examples the space-time is fixed and we only chose different
space-like surfaces on it.
The space-time metric is given in Schwarzschild coordinates $(t,r_s,\theta,\phi)$ by
\begin{equation}
  \label{eq:6}
  ds^2=-\left(1-\frac{2C}{r_s}\right) dt^2+\left(1-\frac{2C}{r_s}\right)^{-1}
  dr_s^2+r_s^2 (d\theta^2+\sin^2\theta d\phi^2).
\end{equation}
These coordinates are singular at $r_s=2C$ and hence they do not reveal the
global structure of the surfaces $t=constant$. The most direct way to see that
these surfaces are complete 3-dimensional manifolds is using the isotropical
radius $r$ defined by
\begin{equation}
  \label{eq:103}
  r_s=r \left(1+\frac{C}{2r} \right)^2.
\end{equation}
In isotropic coordinates the line element is given by
\begin{equation}
  \label{eq:18}
  ds^2=-\left(\frac{1-\frac{C}{2r}}{1+\frac{C}{2r}}\right)^2 dt^2
  +\left(1+\frac{C}{2r}\right)^4(dr^2+d\theta^2+r^2 \sin^2\theta d\phi^2).
\end{equation}
The initial data on the slice $t=constant$ are given by
\begin{equation}
  \label{eq:40b}
  h_{ij}=\left(1+\frac{C}{2r}\right)^4\delta_{ij}, \quad K_{ij}=0.
\end{equation}
These are the time symmetric initial data studied in section
\ref{s:energy}. The linear momentum of these data is obviously zero, then the
total mass $M$ is equal to the energy $E$ calculated in the previous section
and we obtain the expected result $M=C$.

We take another foliation of space-like surfaces.
We write  the metric (\ref{eq:18}) in the Gullstrand -- Painlev\'e coordinates
$(t_{gp},r_s,\theta, \phi)$ (see \cite{Martel:2000rn} and reference therein).  We
obtain
\begin{equation}
  \label{eq:17}
  ds^2=-(1-\frac{2C}{r_s})dt_{gp}^2+2\sqrt{\frac{2C}{r_s}}dt_{gp} dr_s+
dr_s^2+r_s^2 d\theta^2+r_s^2 \sin^2\theta d\phi^2.
\end{equation}
The slices $t_{gp}=constant$ in these coordinates have the following initial
data
\begin{equation}
  \label{eq:11}
  h_{ij}=\delta_{ij}, \quad K_{ij}= \frac{\sqrt{2m}}{ r_s^{3/2}}\left(\delta_{ij}
    -\frac{3}{2}s_is_j \right),
\end{equation}
where $s^i$ is the radial unit normal vector with respect to the flat metric
$\delta_{ij}$.  We see that the intrinsic metric is flat and hence the energy
$E$ is clearly zero. The linear momentum is also zero, because if we calculate
the integral (\ref{eq:EPP}) at an sphere of finite radius (note that the limit
is in danger to diverge because the radial dependence of $K_{ij}$ in
(\ref{eq:11})) the angular variables integrate to zero. And hence we obtain
that for these surfaces the total mass $M$ is zero. What happens is that the
second fundamental form (\ref{eq:11}) does not satisfy the decay condition
(\ref{eq:99}) since it falls off like $O(r^{-3/2})$. It can be proved that any
initial conditions that satisfy (\ref{eq:99}) in the same space-time give the
same total mass $M$.

We consider another foliation which reveal the Lorentz transformation
properties of $(E,P^i)$.  Let $(x,y,z)$ be the associated Cartesian coordinates
of the isotropical coordinates $(r,\theta,\phi)$, that is
\begin{equation}
  \label{eq:108}
  x=r\cos\phi\sin\theta, \quad y=r\sin\phi\cos\theta, \quad z=r\cos\theta.
\end{equation}
We consider the line element (\ref{eq:18}) written in terms of the coordinates
$(t,x,y,z)$ and we perform the following change of coordinates which represents
a boost in the $z$ direction
\begin{align}
  \label{eq:42b}
  \hat t &= \gamma^{-1} ( t-v  z),\\
 \hat z &= \gamma^{-1} (-v t+\hat z),\\
\hat x &=  x,\\
\hat y & =  y,
\end{align}
where $v$ is a constant and $\gamma=\sqrt{1-v^2}$. Consider a surface $\hat t=
constant$ in these coordinates. The intrinsic metric is given by
\begin{equation}
  \label{eq:43}
  h=\psi^4 (d\hat x^2 +d\hat y^2)+\gamma^{-2} ( -N^2 v^2 + \psi^4) d\hat z^2 ,
\end{equation}
where
\begin{equation}
  \label{eq:44}
  \psi=1+\frac{C}{2r},\quad N=\frac{1-\frac{C}{2r}}{1+\frac{C}{2r}}.
\end{equation}
The radius $r$ can be written in terms of the hat coordinates as follows
\begin{equation}
  \label{eq:45}
  r=\sqrt{x^2+y^2+z^2}=\sqrt{\hat x^2+\hat y^2+\gamma^{-2}(v\hat t +\hat z)^2}.
\end{equation}
One can check that the metric $h$ given by (\ref{eq:43}) is asymptotically flat
in the coordinates $(\hat x, \hat y, \hat z)$. Then, we can compute the energy
of this metric and we obtain
\begin{equation}
  \label{eq:46}
  E=\gamma^{-1}C.
\end{equation}
To obtain the linear momentum we need to compute the second fundamental form of
the slice. The calculations are long (see, for example, \cite{gentile10} for
details), the final result is the following
\begin{equation}
  \label{eq:47}
  P_x=0,\quad P_y=0, \quad P_z= v C \gamma^{-1}.
\end{equation}
Using (\ref{eq:46}) and (\ref{eq:47}) we obtain
\begin{equation}
  \label{eq:48}
  M=\sqrt{E^2-P^iP^j \delta_{ij}}=C.
\end{equation}
That is, the quantities $E,P^i$ transform like a 4-vector under asymptotic
Lorentz transformations of coordinates.

\section{Proof}
\label{s:proof}
In section \ref{s:energy} we have presented two proofs of the positive energy
theorem for two particular cases, for
other proofs that applies to other relevant particular cases (like spherical
symmetry and weak field limit) see \cite{brill80}, \cite{chrusciel12},
\cite{chrusciel08} and references therein.

The first general proof of the positive energy theorem was done by Schoen and
Yau \cite{Schoen79b}. Shortly after it was followed by a proof by Witten
\cite{witten81} using completely different methods. The proof of the Penrose
inequality done by Huisken and Illmanen in \cite{Huisken01} (we briefly discuss
this work in the following section \ref{s:gi}) also provide a new proof
of the positive energy theorem (which is based on an idea of  Geroch \cite{geroch73b})

The simpler of all these proofs is, by far, Witten's one. Also it resembles
other positivity proofs in physics: the total energy is written as a positive
definite integral in the space. In this section we review this proof. The aim
is to present all the relevant steps in the most elementary way.

This proof uses, in an essential way, spinors. We refer the reader to the
chapter in this book by Robert Geroch for an introduction to this subject. We
will follow the notation of that chapter in this section.

There exists various reformulations of the original proof by Witten, in this
section we essentially follow references \cite{Penrose86}
\cite{Szabados04} \cite{horowitz82} \cite{Reula84}.

The proof uses only spinors defined on the spacelike surface, however it is
more transparent to begin with spinor fields in the spacetime and then, at the
very end, to restrict them to the spacelike surface.  Also, this way of
constructing the proof easily generalize to the  proof of the positivity of the
energy at null infinity (Bondi mass) (see  \cite{Reula84}).

Let $(M,g_{ab})$ be a four dimensional Lorenzian manifold with connection
$\nabla_a$. In this section we use the signature $(+---)$ to be consistent with
the literature on spinors. Unfortunately this signature gives a negative sign
to the Riemannian metrics on spacelike surfaces used in the previous sections.

Let $\lambda_A$ be an spinor field in the spacetime, the spin connection is
denoted by $\nabla_{AA'}$, and we use the standard notation $a=AA'$ to
identify spinor indices with tensorial indices.

The proof of the positive energy theorem is based on the remarkable properties
of a 2-form $\boldsymbol{\Omega}$ called the Nester-Witten form \cite{witten81}
\cite{Nester:1982tr}, defined as follows.  The computations of this section
involve integration on different kind of surfaces and hence it is convenient to
use differential forms instead of ordinary tensors. We will denote them with
boldface and no indices (for an introduction to forms see, for example, appendix B in
\cite{Wald84}, we will follow the notation and convention of this reference).

Consider the following complex tensor
\begin{equation}
  \label{eq:66}
  \Omega_{ab}= - i \bar\lambda_{B'} \nabla_{AA'} \lambda_{B}.
\end{equation}
From this tensor we construct the complex 2-form $\boldsymbol{\Omega}$ by
\begin{equation}
  \label{eq:67}
  \boldsymbol{\Omega}= \Omega_{[ab]}.
\end{equation}
Explicitly we have
\begin{equation}
  \label{eq:58}
 \boldsymbol{\Omega}= \frac{i}{2} \left(\bar\lambda_{A'} \nabla_{BB'}
   \lambda_{A}- \bar\lambda_{B'} \nabla_{AA'} \lambda_{B} \right).
\end{equation}
The forms used in the following are always tensor fields
(usually complex) but they are constructed out of spinors, as
in the case of $\boldsymbol{\Omega}$. In order to define these
forms we need to antisymmetrize tensorial indices, to avoid a
complicated notation we will always define first the tensor
field in terms of spinors (as in equation \eqref{eq:66}) and
then define the form antisymmetrizing the tensor indices (as in
\eqref{eq:67}). When there are more than two tensorial indices
the explicit expression of the differential form (like
\eqref{eq:58}) can be lengthy and it is not usually needed. The
spinor $\lambda_A$ has an associated (future directed) null
vector $\xi_a$ given by
\begin{equation}
  \label{eq:75}
  \xi^a=\lambda^A\bar\lambda^{A'}.
\end{equation}
Note that the $\boldsymbol{\Omega}$ can not be written in terms of derivatives
of pure tensors fields like $\xi_a$ and $\nabla_a$.

The strategy of the proof is the following. Consider the exterior derivative
$d\boldsymbol{\Omega}$ (which is a 3-form) and integrate it on a spacelike,
asymptotically flat, 3-surface $S$. Using Stoke's theorem we obtain
\begin{equation}
  \label{eq:64}
 \sum_k \lim_{r\to \infty} \oint_{\Su_r}
  \boldsymbol{\Omega} = \int_S d\boldsymbol{\Omega}.
\end{equation}
We are assuming that $S$ is an asymptotically euclidean manifold with $k$
asymptotic ends $U_k$.  The 2-form $\boldsymbol{\Omega}$ has two important
properties.  The first one is that the left hand side of (\ref{eq:64}) gives is
the total energy-momentum of a prescribed asymptotic end. The second is that the
integrand of the right hand side is non-negative. Both properties depend on the
way in which the spinor field $\lambda^A$ is prescribed.

We begin with the first property. Note that the integrand in the left hand side
of \eqref{eq:64} is complex. But the imaginary part
of $\boldsymbol{\Omega}$ is given by
\begin{equation}
  \label{eq:30}
  \boldsymbol{\Omega}-\bar{\boldsymbol{\Omega}}=i\nabla_{[a} \xi_{b]}=i \,
d \boldsymbol{\xi},
\end{equation}
where, to be consistent with our notation, we write $\boldsymbol{\xi}$ for the
the 1-form $\xi_a$. That is, the imaginary part is the exterior derivative of a
1-form and hence its  integral over a closed 2-surface is zero. Hence the
boundary integral is always real, for arbitrary spinors $\lambda^A$.

To prove the desired property, we need to impose fall off conditions on the
spinor $\lambda^A$. Fix one arbitrary end $k$ (from now on we will always work
on that end, and hence we suppress the label $k$). Let
$\mathring{\lambda}^A$ be an arbitrary \emph{constant} spinor, we require that
the spinor $\lambda^A$ satisfies on that end
\begin{equation}
  \label{eq:23}
  \lambda^A=\mathring{\lambda}^A+\gamma^A, \quad \gamma^A=O(r^{-1}).
\end{equation}
We also assume that the partial derivatives of $\gamma^A$ are $O(r^{-2})$ and we
require that $\lambda^A$ \emph{decays to zero at every other end}.

The idea is to prove that at the chosen end we have
\begin{equation}
  \label{eq:24}
  P_a\mathring{\xi}^a=\frac{1}{8\pi}\lim_{r\to \infty} \oint_{\Su_r}
  \boldsymbol{\Omega},
\end{equation}
where $P_a=(E,P_i)$, with  $E$ and $P_i$  defined
by (\ref{eq:EP})--(\ref{eq:EPP}), and $\mathring{\xi}^a$ is the constant null
vector determined by the constant spinor $\mathring{\lambda}^A$ by
\begin{equation}
  \label{eq:75b}
  \mathring{\xi}^a=\mathring{\lambda}^A\mathring{\bar\lambda}^{A'}.
\end{equation}
Note that the boundary integral in the right hand side of \eqref{eq:24}
determines both the energy and the linear momentum of the end.

To prove \eqref{eq:24} the most important step is to prove that the value of the integral
depends only on the constant spinor $\mathring{\lambda}^A$ and not on
$\gamma^A$. We emphasize, as we will see, that a naive counting of the fall
behaviour of the different terms in $\boldsymbol{\Omega}$, under the assumption
\eqref{eq:23}, does not prove this result. Using  the decomposition
(\ref{eq:23})  we write $\boldsymbol{\Omega}$ as
\begin{equation}
  \label{eq:36}
  \boldsymbol{\Omega}=\mathring{\boldsymbol{\Omega}}+\boldsymbol{\Gamma},
\end{equation}
where
\begin{equation}
  \label{eq:50}
 \mathring{\Omega}_{ab}=- i \mathring{\bar\lambda}_{B'} \nabla_{AA'}
 \mathring{\lambda}_{B}, \quad \mathring{\boldsymbol{\Omega}}=\mathring{\Omega}_{[ab]},
\end{equation}
and
\begin{equation}
  \label{eq:51}
  \Gamma_{ab}=-i\left(\mathring{\bar\lambda}_{B'} \nabla_{AA'}
 \gamma_{B}  + \bar\gamma_{B'} \nabla_{AA'}
    \mathring{\lambda}_{B}+   \bar\gamma_{B'} \nabla_{AA'}
    \gamma_{B}   \right), \quad \boldsymbol{\Gamma}=\Gamma_{[ab]}.
\end{equation}
That is, $\mathring{\boldsymbol{\Omega}}$ depends only on
$\mathring{\lambda}^A$.

We would like to prove that $\boldsymbol{\Gamma}=O(r^{-3})$ and hence it does
not contribute to the integral at infinity \eqref{eq:24}. Consider the third
term in \eqref{eq:51}. The covariant derivative $\nabla_{AA'} \gamma_{B}$ has
two terms, the first one contains partial derivatives of $\gamma_{B}$ which, by
assumption, are $O(r^{-2})$. The second term contains products of $\gamma_{B}$
and the connections coefficients of the space-time metric $g_{ab}$ evaluated at
the asymptotic end of the spacelike surface $S$. These coefficients are first
derivatives of $g_{ab}$, they can be written as first derivatives of the
intrinsic Riemannian metric and the second fundamental form of the surfaces and
hence, by assumption (recall that $S$ is asymptotically flat and hence we have
the fall off conditions \eqref{eq:99}) they are $O(r^{-2})$. We conclude that
$\nabla_{AA'} \gamma_{B}=O(r^{-2})$ and hence $\bar\gamma_{B'} \nabla_{AA'}
\gamma_{B} =O(r^{-3})$. We proceed in a similar way for the second term: since
$\mathring{\lambda}^A$ is constant the covariant derivative $\nabla_{AA'}
\mathring{\lambda}_{B}$ contains connection coefficients times constants and
hence we have $\nabla_{AA'}\mathring{\lambda}_{B} =O(r^{-2})$, and then
$\bar\gamma_{B'} \nabla_{AA'} \mathring{\lambda}_{B}=O(r^{-3})$. But using the
same argument we obtain that the first term in \eqref{eq:51} is $O(r^{-2})$ and
then it can contribute to the integral. But we can
re-write  $\Gamma_{ab}$ as follows
\begin{equation}
  \label{eq:91}
  \Gamma_{ab}=-i\left( \nabla_{BB'} (\gamma_A  \mathring {\bar{\lambda}}_{A'})-
\gamma_{A} \nabla_{BB'}
    \mathring{\lambda}_{A} + \bar\gamma_{B'} \nabla_{AA'}
    \mathring{\lambda}_{B}+   \bar\gamma_{B'} \nabla_{AA'}
    \gamma_{B}        \right).
\end{equation}
The first term in (\ref{eq:91}),  which is the problematic one, contribute to
$\boldsymbol{\Gamma}$  with
the derivative of a 1-form, and hence it integrate to zero over a closed
2-surface. The new second term in \eqref{eq:91} is clearly $O(r^{-3})$.
We have proved that
\begin{equation}
  \label{eq:26}
  \lim_{r\to \infty} \oint_{\Su_r}
  \boldsymbol{\Omega}=\lim_{r\to \infty} \oint_{\Su_r}
 \mathring{\boldsymbol{\Omega}}.
\end{equation}
Note that $\mathring{\boldsymbol{\Omega}}$ is $O(r^{-2})$ and hence the
integral converges. Also, the asymptotic value of
$\mathring{\boldsymbol{\Omega}}$ at infinity contain a combination of first
derivative of the intrinsic metric and the second fundamental form of the
surface $S$ multiplied by the constants $\mathring{\lambda}^A$. It can be
proved, essentially by an explicit calculation, that this combination is
precisely $P_a\xi^a$ (see \cite{witten81}, \cite{Nester:1982tr} and also
\cite{Bizon:1986ea}).

We turn to the second property of $\boldsymbol{\Omega}$. Recall that the
exterior derivative of a $p$-form is given by
\begin{equation}
  \label{eq:63}
  d\boldsymbol{\Omega}=(p+1)\nabla_{[a} \Omega_{b_1\cdots b_p]}.
\end{equation}
We have
\begin{equation}
  \label{eq:59}
  d \boldsymbol{\Omega} = \boldsymbol{\alpha} +\boldsymbol{\beta},
\end{equation}
where $\boldsymbol{\alpha}$ and $\boldsymbol{\beta}$ are the following 3-forms
\begin{equation}
  \label{eq:68}
\alpha_{abc} =  -i \bar\lambda_{C'}\nabla_a \nabla_b \lambda_C,\quad
\boldsymbol{\alpha}= \alpha_{[abc]},
\end{equation}
and
\begin{equation}
  \label{eq:60}
  \beta_{abc} =-i \nabla_a\bar\lambda_{C'} \nabla_b \lambda_C\quad
  \boldsymbol{\beta}= \beta_{[abc]} .
\end{equation}
That is, $\boldsymbol{\alpha}$ has second derivatives of the spinor $\lambda_A$
and $\boldsymbol{\beta}$ has squares of first derivatives of $\lambda_A$.

We compute first $\boldsymbol{\alpha}$. Observe that there is a commutator of covariant
derivatives and hence we can replace it by the curvature tensor. However,
what is surprising is that precisely the Einstein tensor appears. To see this, is
easier to work with the dual of $\boldsymbol{\alpha}$ defined by
\begin{equation}
  \label{eq:65}
  {}^*\boldsymbol{\alpha}=\frac{1}{3!}\epsilon_{abcd}\alpha^{abc}.
\end{equation}
We use the conmutator relations
\begin{equation}
  \label{eq:69}
  2\nabla_{[a} \nabla_{b]}\lambda_C=-\epsilon_{A' B'}   X_{ABC} {}^E\lambda_E
  -\epsilon_{A B}   \Phi_{ A' B' C} {}^E\lambda_E,
\end{equation}
where $X_{ABCD}$ and $\Phi_{A' B' CD}$ are the curvature spinors. These spinors
are defined in terms of the Riemann tensor $R_{abcd}=R_{AA' BB'CC"DD"}$ by
\begin{equation}
  \label{eq:84}
  X_{ABCD}=\frac{1}{4} R_{AX'B}{}^{X'}{}_{CY'D}{}^{Y'}, \quad
  \Phi_{ABC'D'}=\frac{1}{4} R_{AX'B}{}^{X'}{}_{YC'}{}^{Y}{}_{D'}. 
\end{equation}
See \cite{Penrose84} for further details on the curvature spinors. 
The Einstein
tensor is given by
\begin{equation}
  \label{eq:70}
  G_{ab}=-6\Lambda g_{ab}-\Phi_{ab},
\end{equation}
where $\Lambda$ is given by
\begin{equation}
  \label{eq:71}
  \Lambda=\frac{1}{6}X_{AB}{}^{AB}.
\end{equation}
We also  use the identities
\begin{equation}
  \label{eq:72}
  X_{ABC}{}^{B}=3\Lambda \epsilon_{AC},
\end{equation}
and
\begin{equation}
  \label{eq:73}
  \epsilon_{abcd}=i\left(\epsilon_{AC}\epsilon_{BD}\epsilon_{A'D'}\epsilon_{B'C'}-\epsilon_{AD}\epsilon_{BC}\epsilon_{A'C'}\epsilon_{B'D'}  \right).
\end{equation}
And then we obtain
\begin{equation}
  \label{eq:74}
   {}^*\boldsymbol{\alpha}=-\frac{1}{2 \cdot 3!} \xi_e G^{ef},
\end{equation}
and hence
\begin{equation}
  \label{eq:61}
 \boldsymbol{\alpha}=-\frac{1}{2 \cdot 3!} \xi_e G^{ef}  \epsilon_{fabc}.
\end{equation}
The expressions \eqref{eq:74} and \eqref{eq:61} are pure tensorial expressions.

To compute $\boldsymbol{\beta}$ we proceed in a similar form. We work first
with the dual
\begin{equation}
  \label{eq:62}
  {}^*\boldsymbol{\beta}=\frac{1}{3!}\epsilon_{abcd}\beta^{bcd}.
\end{equation}
It is important (we see later why) to split the covariant derivative $\nabla_a$
into its temporal and  spatial component. Let $t^a$ denote the unit timelike
normal to the surface $S$ and $h_{ab}$ is the intrinsic metric of the surface.
We define the spatial  $\mathcal{D}_a$ derivative as
\begin{equation}
  \label{eq:78}
  \mathcal{D}_a=h_a{}^b\nabla_b.
\end{equation}
Note that $\mathcal{D}_a$ is not the covariant derivative $D$ of the intrinsic
metric $h$ used in the previous sections, they are related by the equation
\begin{equation}
  \label{eq:27}
  \mathcal{D}_{AB}\lambda_C= D_{AB}\lambda_C + \frac{1}{\sqrt{2}}\pi_{ABC}{}^D \lambda_D,
\end{equation}
where $\pi_{ABCD}=\pi_{(AB)(CD)}$ is the spinor representation of the second
fundamental fundamental form of the surface.

From equation \eqref{eq:78} we obtain
\begin{equation}
  \label{eq:81}
  \nabla_a=\mathcal{D}_a -t_a t^b\nabla_b.
\end{equation}
We replace the derivative $\nabla_a$ by  (\ref{eq:81}) in
the definition of $\boldsymbol{\beta}$ given by  (\ref{eq:60}) and we compute
the dual defined by \eqref{eq:62} to obtain
\begin{equation}
  \label{eq:82}
   {}^*\boldsymbol{\beta}=-i \frac{1}{3!}\epsilon_{abcd} \left(
   \mathcal{D}^b\bar\lambda^{C'} \mathcal{D}^d \lambda^C \right)+ W_a,
\end{equation}
where
\begin{equation}
  \label{eq:29}
  W_a= i \frac{1}{3!}\epsilon_{abcd} \left( t^b t^f\nabla_f
   \bar\lambda^{C'} \mathcal{D}^d \lambda^C +t^d t^f\nabla_f
  \lambda^C \mathcal{D}^b \bar\lambda^{C'}\right).
\end{equation}
Note that $W_a$ satisfies
\begin{equation}
  \label{eq:40}
  t^a W_a=0.
\end{equation}
Using the identity (\ref{eq:73})  we further decompose  the first term in the
right hand side of \eqref{eq:82}
\begin{align}
  \label{eq:83}
 -i \epsilon_{abcd}
   \mathcal{D}^b\bar\lambda^{C'} \mathcal{D}^d \lambda^C &=
   \mathcal{D}^b \bar\lambda^{B'} \mathcal{D}_{BA'}\lambda_{A}
   -\mathcal{D}^{b}\bar\lambda_{A'} \mathcal{D}_{AB'}\lambda_{B}. \\
& = \mathcal{D}_{C'B} \bar \lambda^{C'} \mathcal{D}^{B}{}_{A'} \lambda_{A}
+\mathcal{D}_{CB'}  \lambda^{C} \mathcal{D}^{B'}{}_{A} \bar\lambda_{A'}
-\mathcal{D}_b \lambda_{A} \mathcal{D}^b \bar \lambda_{A'},
\end{align}
where in the second line we have used the spinorial identity
\begin{equation}
  \label{eq:42}
  \epsilon_{AB}\epsilon_{CD}+\epsilon_{BC}\epsilon_{AD} + \epsilon_{CA}\epsilon_{BD}=0.
\end{equation}
Combining \eqref{eq:82} and \eqref{eq:83} we finally obtain
\begin{equation}
  \label{eq:28}
 {}^*\boldsymbol{\beta}=   \mathcal{D}_{C'B} \bar \lambda^{C'} \mathcal{D}^{B}{}_{A'} \lambda_{A}
+\mathcal{D}_{CB'}  \lambda^{C} \mathcal{D}^{B'}{}_{A} \bar\lambda_{A'}
-\mathcal{D}_b \lambda_{A} \mathcal{D}^b \bar \lambda_{A'}  +W_a.
\end{equation}

We are in position now to perform the integral over $S$ of
$d\boldsymbol{\Omega}$.   Using \eqref{eq:59}, \eqref{eq:61} and \eqref{eq:28}
we obtain
\begin{equation}
  \label{eq:49}
  \int_S d\boldsymbol{\Omega}=\int_S \left( 4\pi T_{ab} \xi^b +
\mathcal{D}_{C'B} \bar \lambda^{C'} \mathcal{D}^{B}{}_{A'} \lambda_{A}
+\mathcal{D}_{CB'}  \lambda^{C} \mathcal{D}^{B'}{}_{A} \bar\lambda_{A'}
-\mathcal{D}_b \lambda_{A} \mathcal{D}^b \bar \lambda_{A'}
\right) t^{a}\dv,
\end{equation}
where we have used Einstein equations
\begin{equation}
  \label{eq:57}
  G_{ab}=8\pi T_{ab},
\end{equation}
to replace the Einstein tensor by the energy-momentum tensor in the expression
\eqref{eq:61} for $\boldsymbol{\alpha}$. Note that the term $W_a$ in
\eqref{eq:28}  does not appear in the integral because it is orthogonal to
$t^a$ (c.f. equation \eqref{eq:40}).

Assume that the spinor $\lambda^A$ has the fall-off behaviour \eqref{eq:23},
then the identity \eqref{eq:24} holds, using  the Stoke's theorem
\eqref{eq:64} (note that by \eqref{eq:23} all the other boundary integrals
vanish) we finally obtain the famous Witten identity
\begin{equation}
  \label{eq:109}
  P_a\mathring{\xi}^a= \frac{1}{8\pi}\int_S \left( 4\pi T_{ab} \xi^b +
\mathcal{D}_{C'B} \bar \lambda^{C'} \mathcal{D}^{B}{}_{A'} \lambda_{A}
+\mathcal{D}_{CB'}  \lambda^{C} \mathcal{D}^{B'}{}_{A} \bar\lambda_{A'}
-\mathcal{D}_b \lambda_{A} \mathcal{D}^b \bar \lambda_{A'}
\right) t^{a}\dv,
\end{equation}
If we assume that the energy-momentum tensor $T_{ab}$ satisfies the dominant
energy condition then we have
\begin{equation}
  \label{eq:110}
  T_{ab} \xi^a t^b\geq 0,
\end{equation}
and hence the first term in the integrand of \eqref{eq:109} is
non-negative. The last term in \eqref{eq:109} is also non-negative since it involves the
contraction with the Riemannian metric (which is negative definite) and the
timelike  vector $t^{AA'}$.
To handle the second and third term we impose on
$\lambda^A$ the following equation which is called the Sen-Witten equation
\cite{sen81} \cite{witten81}
\begin{equation}
  \label{eq:79}
   \mathcal{D}_{AB}\lambda^A=0.
\end{equation}
Let us assume for the moment that there is a solution of this equation with the
fall-off behaviour \eqref{eq:23}.
Then, from \eqref{eq:109} we obtain
\begin{equation}
  \label{eq:111}
  P_a\mathring{\xi}^a\geq 0.
\end{equation}
But the constant null vector $\mathring{\xi}^a$ is arbitrary, hence it follows that $P_a$ should
satisfy \eqref{eq:100}.
To prove the rigidity part of theorem \ref{t:pmt} the key ingredient is that
$E=0$ implies, again by the identity \eqref{eq:109}, that the spinor satisfies
the equation
\begin{equation}
  \label{eq:41}
  \mathcal{D}_{AB}\lambda_C=0,
\end{equation}
that is, it is covariant constant in the whole manifold. From this equation
it can be deduced that the initial data on the surface correspond to the
Minkowski space-time (see \cite{Parker82} for the details of this argument).

It remains to discuss the solutions of equation \eqref{eq:79}. The existence of
solution of this equations under the required fall-off conditions \eqref{eq:23}
has been proved in \cite{Reula82} \cite{Reula84} \cite{Parker82}. The main
point is that equation \eqref{eq:79} constitute an elliptic system of first
order for the two complex components of the spinor (this can be easily seen
using the standard definition of ellipticity for systems, see, for example,
\cite{Dain06} where this specific example is discussed). And hence this
equation can, essentially, be handled as a Poisson equation. Solutions under
weak decay conditions on the data of equation \eqref{eq:79} has been proved in
\cite{Bizon:1986ea}.

Finally, let us discuss the proof of theorem \ref{t:pmtbh}. This was done in
\cite{Reula84} \cite{Gibbons83}.  Remarkably, the proof is very
similar, the only extra ingredient is that in the Stoke's theorem we need to
include an extra internal boundary term. This term has the form (see \cite{Reula84})
\begin{equation}
  \label{eq:76}
  \int_{\partial B} \boldsymbol{\Omega}= \int_{\partial B} \left( \Theta_{+}
   \lambda^0\bar \lambda^{0'} -\rho'  \lambda^1\bar \lambda^{1'}
+\lambda^{1'}  \eth  \lambda^0 - \bar  {\lambda}^0 \bar\eth \lambda^1\right)\, ds.
\end{equation}
In this equation $\Theta_{+}$ is the null expansion defined previously by
\eqref{eq:96}. The coefficient $\rho'$ represent the ingoing null expansion on
the surface, it is not important for our purposes. The functions $\lambda^0$
and $\lambda^1$ are the component of $\lambda^A$ in an appropriated spinorial
diad adapted to the 2-surface $\partial B$. Finally, $\eth$ is a tangential
differential operator to the 2-surface. It can be shown that the appropriate
inner Dirichlet boundary for equation \eqref{eq:79} is to prescribe one of the
component $\lambda^0$ or $\lambda^1$ (but not both) (this is a consequence of
the elliptic character of this equation, see, for example \cite{Dain06} for an
elementary treatment of this).  If we prescribe $\lambda^1=0$ on $\partial B$
and use that, by hypothesis this surface satisfies $\Theta_{+}=0$, then the
boundary term \eqref{eq:76} vanished and we can proceed in the same way as
above to prove the positivity of the energy. Note that without the condition
$\Theta_{+}=0$ it is not possible to make the boundary term zero.

\section{Further results and open problems}
\label{s:gi}
In this article we have discussed only the positive energy theorem in 3 space
dimensions. The spinorial proof presented in section \ref{s:proof} works in any
dimensions (see \cite{Parker82}), however in higher dimensions the existence of
an spin structure involves restrictions on the topology of the manifold $S$. In
section \ref{s:proof} we have used Weyl spinors which are well adapted to 4
spacetime dimensions. For higher dimensions Dirac spinors are usually used. The
other proofs currently available \cite{Schoen79b} and \cite{Huisken01} do not
work in arbitrary high dimensions. To prove the positive energy theorem in all
dimensions is one of the relevant open problem in this area.

The positive energy theorem can be refined to incorporate other physically
relevant parameters.  For example, using a similar argument as in Witten's
proof it is possible to prove \cite{Gibbons83} \cite{Gibbons82}  that the total mass $M$
satisfies
\begin{equation}
  \label{eq:77}
  M\geq |q|,
\end{equation}
where $q$ is the electric charge and the non-electromagnetic part of the energy
momentum tensor must satisfies appropriated conditions.

Recently, for axially symmetric black holes  the following inequality
\begin{equation}
    \label{eq:53}
    M \geq \sqrt{|J|},
  \end{equation}
  has been proved. Here $J$ is the angular momentum of the black hole (see the
  review article \cite{dain12} and reference therein). The equality in
  \eqref{eq:53} is achieved only for the extreme Kerr black hole. This
  inequality is proved for one black hole, a relevant open problem is to
  prove it for multiple black holes. 

Another important extension is the Penrose inequality for black holes. The
Riemannian black hole positivity theorem \ref{t:bhrpm} can be generalized to
include the area of the minimal surface, namely
  \begin{equation}
    \label{eq:52}
    M\geq \sqrt{\frac{A}{16\pi}},
  \end{equation}
with equality only for the Schwarzschild black hole. This result was proved by
\cite{Huisken01} and \cite{Bray01}. The general case remains open,
see the review article \cite{Mars:2009cj}.

Finally, we have discussed the concept of total energy and linear momentum of
an isolated system. It would be very desirable to have a quantity that measure
the energy of a finite region of the spacetime. These kind of quantities are
called quasi-local mass. For a comprehensive review on this important open
problem see \cite{Szabados04}. The following related, pure quasi-local, inequality for
axially symmetric black holes has been recently proved
\begin{equation}
  \label{eq:85}
  A\geq 8\pi |J|,
\end{equation}
where $A$ is the area and $J$ is the quasi-local angular momentum of the black
hole (see the review article \cite{dain12} and reference therein).  The
equality in (\ref{eq:85}) is achieved if and only if the local geometry of the
black hole is equal to the extreme Kerr black hole local geometry.  For
non-axially symmetric black holes it is difficult to define the quasi-local
angular momentum $J$ (see \cite{Szabados04}). An important open problem is to
generalize the inequality (\ref{eq:85}) for non-axially symmetric black holes
(or to find suitable counter examples).


\begin{thebibliography}{10}

\bibitem{alcubierre-it3nrsomop2008}
M.~Alcubierre.
\newblock {\em Introduction to {$3+1$} numerical relativity}, volume 140 of
  {\em International Series of Monographs on Physics}.
\newblock Oxford University Press, Oxford, 2008.

\bibitem{Arnowitt62}
R.~Arnowitt, S.~Deser, and C.~W. Misner.
\newblock The dynamics of general relativity.
\newblock In L.~Witten, editor, {\em Gravitation: An Introduction to Current
  Research}, pages 227--265. Wiley, New York, 1962, gr-qc/0405109.

\bibitem{Ashtekar-horowitz82}
A.~Ashtekar and G.~T. Horowitz.
\newblock Energy-momentum of isolated systems cannot be null.
\newblock {\em Physics Letters A}, 89(4):181--184, 1982.

\bibitem{Bartnik86}
R.~Bartnik.
\newblock The mass of an asymptotically flat manifold.
\newblock {\em Comm. Pure App. Math.}, 39(5):661--693, 1986.

\bibitem{Beig96d}
R.~Beig and P.~T. Chru{\'s}ciel.
\newblock {K}illing vectors in asymptotically flat space--times: I.
  asymptotically translational {K}illing vectors and the rigid positive energy
  theorem.
\newblock {\em J. Math. Phys.}, 37:1939--1961, 1996, gr-qc/9510015.

\bibitem{Bizon:1986ea}
P.~Bizon and E.~Malec.
\newblock {On Witten's positive energy proof for weakly asymptotically flat
  space-times}.
\newblock {\em Class.Quant.Grav.}, 3:L123, 1986.

\bibitem{Bray01}
H.~L. Bray.
\newblock Proof of the riemannian penrose conjecture using the positive mass
  theorem.
\newblock {\em J. Differential Geometry}, 59:177--267, 2001, math.DG/9911173.

\bibitem{Bray04}
H.~L. Bray and P.~T. Chru{\'s}ciel.
\newblock The {P}enrose inequality.
\newblock In {\em The Einstein equations and the large scale behavior of
  gravitational fields}, pages 39--70. Birkh\"auser, Basel, 2004.

\bibitem{Bray:2009cw}
H.~L. Bray and J.~L. Jauregui.
\newblock {A Geometric theory of zero area singularities in general
  relativity}, 2009, 0909.0522.

\bibitem{Brill59}
D.~Brill.
\newblock On the positive definite mass of the {B}ondi-{W}eber-{W}heeler
  time-symmetric gravitational waves.
\newblock {\em Ann. Phys.}, 7:466--483, 1959.

\bibitem{brill80}
D.~R. Brill and P.~S. Jang.
\newblock The positive mass conjecture.
\newblock In {\em General relativity and gravitation, {V}ol. 1}, pages
  173--193. Plenum, New York, 1980.

\bibitem{Brill63}
D.~R. Brill and R.~W. Lindquist.
\newblock Interaction energy in geometrostatics.
\newblock {\em Phys. Rev.}, 131:471--476, 1963.

\bibitem{Cantor81b}
M.~Cantor and D.~Brill.
\newblock The {L}aplacian on asymptotically flat manifolds and the
  specification of scalar curvature.
\newblock {\em Compositio Mathematica}, 43(3):317--330, 1981.

\bibitem{chrusciel86}
P.~Chru{\'s}ciel.
\newblock Boundary conditions at spatial infinity from a {H}amiltonian point of
  view.
\newblock In {\em Topological properties and global structure of space-time
  ({E}rice, 1985)}, volume 138 of {\em NATO Adv. Sci. Inst. Ser. B Phys.},
  pages 49--59. Plenum, New York, 1986.

\bibitem{chrusciel08}
P.~T. Chru{\'s}ciel.
\newblock Lectures on mathemacital {R}elativity, 2008.
\newblock
  \url{http://homepage.univie.ac.at/piotr.chrusciel/papers/BeijingAll.pdf}.

\bibitem{chrusciel12}
P.~T. Chru{\'s}ciel.
\newblock Lectures on energy in {G}eneral {R}elativity, 2012.
\newblock \url{http://homepage.univie.ac.at/piotr.chrusciel}.

\bibitem{Cook00}
G.~B. Cook.
\newblock Initial data for numerical {R}elativity.
\newblock {\em Living Rev. Relativity}, 3(5):2000--5, 53 pp. (electronic),
  2001.
\newblock http://www.livingreviews.org/Articles/Volume3/2000-5cook/.

\bibitem{Dain06}
S.~Dain.
\newblock Elliptic systems.
\newblock In J.~Frauendiener, D.~Giulini, and V.~Perlick, editors, {\em
  Analytical and Numerical Approaches to Mathematical Relativity}, volume 692
  of {\em Lecture Notes in Physics}, pages 117--139. Springer, 2006,
  gr-qc/0411081.

\bibitem{dain12}
S.~Dain.
\newblock Geometric inequalities for axially symmetric black holes.
\newblock {\em Classical and Quantum Gravity}, 29(7):073001, 2012, 1111.3615.

\bibitem{Dain:2008yu}
S.~Dain and M.~E. Gabach~Cl\'ement.
\newblock {Extreme Bowen-York initial data}.
\newblock {\em Class. Quantum. Grav.}, 26:035020, 2009, 0806.2180.

\bibitem{denisov83}
V.~I. Denisov and V.~O. Solov'ev.
\newblock The energy determined in general relativity on the basis of the
  traditional {H}amiltonian approach does not have physical meaning.
\newblock {\em Theoretical and Mathematical Physics}, 56:832--841, 1983.
\newblock 10.1007/BF01016826.

\bibitem{gentile10}
I.~Gentile~de Austria.
\newblock Superficies maximales con momento lineal en {S}chwarzschild.
\newblock Master's thesis, Facultad de Matem\'atica Astronom\'ia y F\'isica,
  Universidad Nacional de C\'ordoba, Argentina, 2010.
\newblock \url{http://www.famaf.unc.edu.ar/~dain/ivan-tf.pdf}.

\bibitem{geroch73b}
R.~Geroch.
\newblock Energy extraccion.
\newblock {\em Ann. New York Acad. Sci}, 224:108--117, 1973.

\bibitem{Gibbons83}
G.~W. Gibbons, S.~W. Hawking, G.~T. Horowitz, and M.~J. Perry.
\newblock Positive mass theorems for black holes.
\newblock {\em Commun. Math. Phys.}, 88:295--308, 1983.

\bibitem{Gibbons82}
G.~W. Gibbons and C.~M. Hull.
\newblock A {B}ogomolny bound for general relativity and solitons in {$N=2$}\
  supergravity.
\newblock {\em Phys. Lett. B}, 109(3):190--194, 1982.

\bibitem{horowitz82}
G.~T. Horowitz and P.~Tod.
\newblock A relation between local and total energy in general relativity.
\newblock {\em Communications in Mathematical Physics}, 85:429--447, 1982.
\newblock 10.1007/BF01208723.

\bibitem{Huisken01}
G.~Huisken and T.~Ilmanen.
\newblock The inverse mean curvature flow and the {R}iemannian {P}enrose
  inequality.
\newblock {\em J. Differential Geometry}, 59:352--437, 2001.

\bibitem{Lee87}
J.~M. Lee and T.~H. Parker.
\newblock The {Y}amabe problem.
\newblock {\em Bull. Amer. Math. Soc.}, 17(1):37--91, 1987.

\bibitem{Mars:2009cj}
M.~Mars.
\newblock {Present status of the Penrose inequality}.
\newblock {\em Class. Quant. Grav.}, 26:193001, 2009, 0906.5566.

\bibitem{Martel:2000rn}
K.~Martel and E.~Poisson.
\newblock {Regular coordinate systems for Schwarzschild and other spherical
  space-times}.
\newblock {\em Am.J.Phys.}, 69:476--480, 2001, gr-qc/0001069.

\bibitem{Misner60}
C.~W. Misner.
\newblock Wormhole initial conditions.
\newblock {\em Phys. Rev.}, 118:1110--1111, 1960.

\bibitem{murchadha:2111}
N.~O. Murchadha.
\newblock Total energy momentum in {G}eneral {R}elativity.
\newblock {\em Journal of Mathematical Physics}, 27(8):2111--2128, 1986.

\bibitem{Murchadha74}
N.~O. Murchadha and J.~W. York.
\newblock {Gravitational energy}.
\newblock {\em Phys. Rev. D}, 10:2345--2357, 1974.

\bibitem{Nester:1982tr}
J.~A. Nester.
\newblock {A New gravitational energy expression with a simple positivity
  proof}.
\newblock {\em Phys.Lett.}, A83:241, 1981.

\bibitem{osserman86}
R.~Osserman.
\newblock {\em A survey of minimal surfaces}.
\newblock Dover Publications Inc., New York, second edition, 1986.

\bibitem{Parker82}
T.~Parker and C.~H. Taubes.
\newblock On {W}itten's proof of the positive energy theorem.
\newblock {\em Commun. Math. Phys.}, 84(2):223--238, 1982.

\bibitem{Penrose84}
R.~Penrose and W.~Rindler.
\newblock {\em Spinors and Space-Time}, volume~1.
\newblock Cambridge University Press, Cambridge, 1984.

\bibitem{Penrose86}
R.~Penrose and W.~Rindler.
\newblock {\em Spinors and Space-Time}, volume~2.
\newblock Cambridge University Press, Cambridge, 1986.

\bibitem{Reula82}
O.~Reula.
\newblock Existence theorem for solutions of {W}itten's equation and
  nonnegativity of total mass.
\newblock {\em J. Math. Phys.}, 23(5):810--814, 1982.

\bibitem{Reula84}
O.~Reula and K.~P. Tod.
\newblock Positivity of the {B}ondi energy.
\newblock {\em J. Math. Phys.}, 25(4):1004--1008, 1984.

\bibitem{Schoen79b}
R.~Schoen and S.~T. Yau.
\newblock On the proof of the positive mass conjecture in general relativity.
\newblock {\em Comm. Math. Phys.}, 65(1):45--76, 1979.

\bibitem{sen81}
A.~Sen.
\newblock On the existence of neutrino ''zero-modes'' in vacuum spacetimes.
\newblock {\em Journal of Mathematical Physics}, 22(8):1781--1786, 1981.

\bibitem{Szabados04}
L.~B. Szabados.
\newblock Quasi-local energy-momentum and angular momentum in {GR}: A review
  article.
\newblock {\em Living Rev. Relativity}, 7(4), 2004.
\newblock cited on 8 August 2005.

\bibitem{Wald84}
R.~M. Wald.
\newblock {\em General Relativity}.
\newblock The University of Chicago Press, Chicago, 1984.

\bibitem{witten81}
E.~Witten.
\newblock A new proof of the positive energy theorem.
\newblock {\em Communications in Mathematical Physics}, 80:381--402, 1981.
\newblock 10.1007/BF01208277.

\end{thebibliography}

\end{document}